\documentclass[12pt,a4paper]{article}
\usepackage{graphicx}

\usepackage{amsmath,amssymb,latexsym}
\usepackage{paralist,array,url}

\usepackage{amsmath}
\usepackage{amssymb}
\usepackage{amsthm}
\usepackage[usenames,dvipsnames]{xcolor}
\usepackage{multirow,url}
\usepackage{alltt}
\usepackage{paralist}
\usepackage{threeparttable}
\usepackage{pifont}

\usepackage{tabulary}

\usepackage[ruled,lined,linesnumbered,boxed]{algorithm2e}

\usepackage{afterpage}

\newcommand{\set}[1]{\left\{#1\right\}}

\begin{document}
\newcommand{\up}{\vspace*{-0.05cm}}
\newcommand{\pf}{\noindent{\bf Proof: }}
\newtheorem{thm}{Theorem}
\newtheorem{lemma}{Lemma}
\newtheorem{prop}{Proposition}
\newtheorem{prob}{Problem}
\newtheorem{quest}{Question}
\newtheorem{ex}{Example}
\newtheorem{cor}{Corollary}
\newtheorem{conj}{Conjecture}
\newtheorem{cl}{Claim}
\newtheorem{df}{Definition}
\newtheorem{rem}{Remark}
\newtheorem{instance}{Instance}
\newtheorem{example}{Example}
\newcommand{\beq}{\begin{equation}}
\newcommand{\eeq}{\end{equation}}
\newcommand{\<}[1]{\left\langle{#1}\right\rangle}
\newcommand{\be}{\beta}
\newcommand{\ee}{\end{enumerate}}
\newcommand{\Bul}{\mbox{$\bullet$ } }
\newcommand{\al}{\alpha}
\newcommand{\ep}{\epsilon}
\newcommand{\si}{\sigma}
\newcommand{\om}{\omega}
\newcommand{\la}{\lambda}
\newcommand{\La}{\Lambda}
\newcommand{\Ga}{\Gamma}
\newcommand{\ga}{\gamma}
\newcommand{\im}{\Rightarrow}
\newcommand{\2}{\vspace{.2cm}}
\newcommand{\es}{\emptyset}
\newcommand{\tick}{\ding{52}}

\title{\Large\bf Algorithms for the workflow satisfiability problem engineered for counting constraints}
\author{
D. Cohen\footnote{e-mail: {\tt D.Cohen@rhul.ac.uk}},
J. Crampton\footnote{e-mail: {\tt Jason.Crampton@rhul.ac.uk}}, 
A. Gagarin\footnote{e-mail: {\tt Andrei.Gagarin@rhul.ac.uk}},
G. Gutin\footnote{e-mail: {\tt G.Gutin@rhul.ac.uk}},
M. Jones\footnote{e-mail: {\tt M.E.L.Jones@rhul.ac.uk}}\\ 
{\footnotesize Royal Holloway, University of London, Egham, Surrey, TW20 0EX, UK}
}
\maketitle

\date{}

\begin{abstract}
The workflow satisfiability problem (WSP) asks whether there exists an assignment of authorized users to the steps in a workflow specification
that satisfies the constraints in the specification. 
The problem is NP-hard in general, but several subclasses of the problem are known to be fixed-parameter tractable (FPT) when parameterized by the number of steps in the specification.  In this paper, we consider the WSP with user-independent counting constraints, a large class of constraints for which the WSP is known to be FPT. We describe an efficient implementation of an FPT algorithm for solving this subclass of the WSP and an experimental evaluation of this algorithm.  The algorithm iteratively generates all equivalence classes of possible partial solutions until, whenever possible, it finds a complete solution to the problem.  We also provide a reduction from a WSP instance to a pseudo-Boolean SAT instance.  We apply this reduction to the instances used in our experiments and solve the resulting PB SAT problems using SAT4J, a PB SAT solver.  We compare the performance of our algorithm with that of SAT4J and discuss which of the two approaches would be more effective in practice.
%
\bigskip


\end{abstract}

\bigskip\bigskip
\section{Introduction}\label{sec:intro}

It is increasingly common for organizations to computerize their business and management processes.
The co-ordination of the steps that comprise a computerized business process is managed by a workflow management system.
Typically, the execution of these steps will be triggered by a human user, or a software agent acting under the control of a human user, and the execution of each step will be restricted to some set of authorized users or agents.
In addition, one may wish to constrain the users who execute certain sets of steps, even if authorized.
We may, for example, require that two particular steps are executed by two different users, in order to enforce some separation-of-duty requirement; or by the same user, to respect a binding-of-duty requirement.

We model a workflow as follows.
We have a set of \emph{steps} $S$, each of which must be performed by some \emph{user} in a set $U$ of users.
We restrict the users that can perform each step with a set of \emph{authorization lists}, 
$\mathcal{A} = \set{A(u) : u \in U}$,
where $A(u) \subseteq S$ denotes the set of steps that user $u$ is authorized to perform.
Furthermore we must also satisfy a set $C$ of \emph{ (business) constraints}.
In general, a \emph{constraint} can be described as a pair $c = (T, \Theta)$, where $T \subseteq S$ and $\Theta$ is a set of functions from $T$ to $U$: $T$ is the \emph{scope} of the constraint and $\Theta$ specifies those assignments of steps in $T$ to users in $U$ that satisfy the constraint.

Given a \emph{workflow} $W = (S,U,\mathcal{A},C)$, $W$ is said to be \emph{satisfiable} if there exists a function $\pi: S \rightarrow U$ such that
\begin{enumerate}
 \item for all $s \in S$, $s \in A(\pi(s))$ (each step is allocated to an authorized user);
 \item for all $(T,\Theta) \in C$, $\pi|_{T} \in \Theta$ (every constraint is satisfied).
\end{enumerate}
Such a function $\pi: S \rightarrow U$ is called a \emph{valid complete plan}.
Evidently, it is possible to specify a workflow that is not satisfiable.
Hence, it is important to be able to determine whether a workflow is satisfiable or not.
This is called the \emph{workflow satisfiability problem} (WSP). 
This problem has been studied extensively in the security research community~\cite{BeFeAt99,Cr05,WaLi10} and more recently as an interesting algorithmic problem~\cite{CoCrGaGuJo13,CrGuYe13}.

 As an example, consider the following instance of the WSP introoduced in \cite{CoCrGaGuJo13}.
 
 \begin{instance} \label{inst:mainExample}
 The task set $S = \{s_1,\dots, s_4\}$ and the user set $U = \{u_1, \dots, u_6\}$.
 The authorization lists are as follows (where a tick indicates that the given user is authorized for the given task):
 
 \begin{center}
    \begin{tabular}{|c | l | l | l | l | l | l |}
    \hline
    & $u_1$ & $u_2$ & $u_3$ & $u_4$ & $u_5$ & $u_6$ \\ \hline
    $s_1$ &\tick &\tick & & & & \\ \hline
    $s_2$ & & \tick& \tick& & & \\ \hline
    $s_3$ & & \tick& &\tick &\tick & \tick \\ \hline
    $s_4$ & & \tick& &\tick &\tick & \tick \\ \hline
    \end{tabular}
\end{center}

The constraints are $(s_1,s_2,=)$, $(s_2,s_3,\neq)$, $(s_3,s_4,\neq)$, and $(s_1,s_4,\neq)$, where $(s_i,s_j,=)$ means that $s_i$ and $s_j$ must be assigned to the same user and $(s_i,s_j,\neq)$ means that $s_i$ and $s_j$ must be assigned to different users.
 
 \end{instance}
 
A function $\pi: T \rightarrow Y$, where $T\subseteq S$ and $Y\subseteq U$, is called a {\em partial plan}. A partial plan $\pi$ is \emph{authorized} if $s\in A(\pi(s))$ for every $s\in T$. A partial plan $\pi$ is \emph{eligible} if it does not violate any constraint in $C$, and $\pi$ is \emph{valid} if it is both authorized and eligible. In other words, a valid partial plan 
could, in principle, be
extended to a valid complete plan.

 Example~\ref{ex:defs} illustrates the meanings of eligible, complete and authorised plans in the context of Instance~\ref{inst:mainExample}.
 
 \begin{example}
 \label{ex:defs}
The following table gives assignments for four plans, $\pi_1, \pi_2, \pi_3, \pi_4$:

 \begin{center}
    \begin{tabular}{|c | l | l | l | l | l | l | l |}
    \hline
    & $s_1$ & $s_2$ & $s_3$ & $s_4$ & Authorized & Eligible & Complete \\ \hline
    $\pi_1$ & $u_1$ &$u_2$ &$u_4$ &$u_5$ &\tick & &\tick \\ \hline
    $\pi_2$ & $u_1$&$u_1$ &$u_4$ &$u_5$ & &\tick &\tick \\ \hline
    $\pi_3$ & $u_1$& - &$u_4$ &$u_5$ &\tick &\tick & \\ \hline
    $\pi_4$ & $u_2$& $u_2$ &$u_4$ &$u_5$ &\tick &\tick &\tick \\ \hline
    \end{tabular}
\end{center}
  
\begin{itemize}
  \item $\pi_1$ is a complete plan which is authorized but not eligible, as $s_1$ and $s_2$ are assigned to different users.
 
  \item $\pi_2$ is a complete plan which is eligible but not authorized, as $u_1$ is not authorized for $s_2$.
  
   \item $\pi_3$ is a plan which is authorized and eligible, and therefore valid.
   However, $\pi_3$ is not a complete plan as there is no assignment for $s_2$.
   
   \item $\pi_4$ is a complete plan which is eligible and authorized. Thus $\pi_4$ is a valid complete plan, and is therefore a solution.
   \end{itemize}
 \end{example}

The WSP is known to be NP-hard~\cite{WaLi10} in general: it is easily shown to be NP-hard even if restricted to simple separation-of-duty constraints. Wang and Li~\cite{WaLi10} observed that, in practice, the number $k$ of steps is usually significantly smaller than the number $n$ of users and, thus, suggested parameterizing the WSP by the number of steps $k$. Also, they showed that, in general, the WSP is W[1]-hard, but is {\em fixed-parameter tractable} (FPT) for certain classes of constraints. 
In other words, for some classes of constraints, the WSP can be solved in time $O^*(f(k))$, where $f$ is an arbitrary function of $k$ only, and $O^*$ suppresses not only constants, but also polynomial factors.
Such algorithms are called {\em FPT}.
For further terminology on parameterized algorithms and complexity, see monographs \cite{DowneyFellows99,FlumGrohe06,Niedermeier06}.

Since the problem is intractable in its generality and covers a vast number of different types of constraints, it is natural to restrict attention to some WSP subclasses defined by the types and properties of constraints.
Many business rules are not concerned with the identities of the users that complete a set of steps. Accordingly, we say a constraint $c = (T,\Theta)$ is \emph{user-independent} if, whenever $\theta \in \Theta$ and $\phi: U \rightarrow U$ is a permutation, then $\phi \theta \in \Theta$. 
In other words, given a plan $\pi$ that satisfies $c$ and any permutation $\phi: U \rightarrow U$, the plan $\pi' : S \rightarrow U$, where $\pi'(s) = \phi(\pi(s))$, also satisfies $c$.

The most obvious example of a user-independent constraint is the requirement that two steps are performed by different users or, in other words, by exactly two users (separation-of-duty). A more complex example might require that at least/at most/exactly $r$ users are required to complete some sensitive set of steps (cardinality or \emph{counting} constraints), where $r$ is usually small, 
normally less than $5$. A constraint that a particular user $u$ has to perform at least three steps, is not user-independent.

There is a substantial literature on constraints as a method for specifying and enforcing business rules (see \cite{GlGaFe98}, for example), including work by researchers at SAP and IBM (see \cite{BaBuKa14,WoSch07}, for example). The most widely studied constraints are counting and separation-of-duty constraints, which form part of the ANSI standard on role-based access control (RBAC) \cite{ANSI04}, developed by the US National Institute of Standards and Technology (NIST). In short, the literature and relevant standards suggest that user-independent constraints are the constraints of most interest in business processing and workflow management systems. 
In particular, all the constraints defined in the ANSI RBAC standard are user-independent.

\subsection{FPT results and analysis}
One of the motivations for this research was to show that the generic FPT algorithm of Cohen {\em et al.} \cite{CoCrGaGuJo13} is not merely of theoretical interest. 
The gap between traditional ``pen-and-paper'' algorithmics and actually implemented computer-feasible algorithms can be enormous~\cite{ChKl10,MyKo11}. 
In this paper, we demonstrate that the generic FPT algorithm of \cite{CoCrGaGuJo13} has practical value and its implementations are able to outperform the well-known PB SAT solver SAT4J \cite{BePa10}.

Crampton {\em et al.}
\cite{CrGuYe13} extended the FPT classes of \cite{WaLi10}  and obtained more efficient algorithms than in \cite{WaLi10}.
Recently, Cohen {\em et al.} \cite{CoCrGaGuJo13} described a new generic algorithm to solve some classes of the WSP. In particular, they proved that their generic algorithm is FPT
for the WSP restricted to the class of user-independent constraints.
Almost all constraints studied in \cite{CrGuYe13,WaLi10} and other papers are user-independent. Since separation-of-duty constraints are user-independent, the WSP restricted to the class of user-independent constraints remains NP-hard \cite{WaLi10}.

In this paper we present two different approaches to solve the WSP with user-independent counting constraints, describe their implementations, and compare their experimental outcomes.  
First, we describe an adaptation of the general FPT algorithm of \cite{CoCrGaGuJo13} to the case of the WSP with user-independent constraints and develop its implementation in the case of counting constraints. 
We then describe a reduction of WSP instances with user-independent counting constraints to a pseudo-Boolean (PB) SAT problem and prove its correctness. This solution approach is similar to the one presented in \cite{WaLi10}.
In this approach, a PB SAT solver is used as a black box 
solver for our WSP instances.

We compare the performance of the two approaches in a set of computational experiments.
While Wang and Li \cite{WaLi10} provided experimental evaluation for their reduction to PB SAT, they did not provide any experimental evaluation for their FPT algorithm for the WSP. 
Our paper therefore represents the first experimental evaluation of an FPT algorithm designed specifically for the WSP.

Our results show that for more challenging well-constrained WSP instances, the FPT algorithm of \cite{CoCrGaGuJo13} is more effective and efficient than the reduction to a PB SAT problem.
In fact, the PB SAT solver (SAT4J) was unable to solve several WSP instances,
usually because of excessive memory requirements.
On the other hand, for lightly-constrained WSP instances, the PB SAT solver  usually outperforms our implementations of the FPT algorithm.

\subsection{Paper organization}

The paper is organized as follows. In Section \ref{sec:solving-wsp:fpt-algorithm}, we describe our generic FPT algorithm and, in Section \ref{sec:implementation}, we describe and discuss its implementation.
In Section \ref{sec:solving-wsp:pb-boolean}, we describe how the family of WSP instances we consider can be formulated as a pseudo-Boolean SAT problem. 
Section \ref{sec:experiments} describes test experiments which we have conducted with synthetic data to 
compare 
our implementations of the FPT algorithm 
to SAT4J.
Finally, Section \ref{sec:conclusion} provides conclusions and discusses plans for future work.

The main differences between this paper and the preliminary version \cite{FAW2014} are as follows. 
In Sections~\ref{sec:solving-wsp:fpt-algorithm}, the generic algorithm for solving user-independent constraints is described.
In Section  \ref{sec:implementation}, its implementation for counting constraints is explained, with a formal proof in Theorem~\ref{prop:useless}  that some users can be skipped during the iteration of the algorithm.
A new heuristic speed-up for the FPT algorithm using pairs of intersecting constraints is described in Subsection~\ref{subsec:Pairs}.
We have also conducted a new set of experimental tests, 
using a new implementation of the FPT algorithm (Section~\ref{sec:experiments}).


\section{Generic FPT algorithm for the WSP}\label{sec:solving-wsp:fpt-algorithm}

In this section we describe how the FPT algorithm of \cite{CoCrGaGuJo13} works in the WSP case with user-independent constraints.
As mentioned above, one of the distinctive features of the WSP is that the number $k=|S|$ of steps is significantly smaller than the number $n=|U|$ of users, which allows us to design efficient FPT algorithms using $k$ as a parameter.
The algorithm presented here iteratively considers users one-by-one and gradually generates 
all possible partial solutions, where two solutions are treated as identical if they satisfy a certain equivalence relation, defined below. The algorithm continues
until it finds a valid complete plan, or all the users have been considered.
Before we give an overview of the algorithm, we introduce some definitions.


For user-independent constraints, two partial plans $\pi : T \rightarrow Y$ and $\pi' : T' \rightarrow Y'$ are \emph{equivalent}, denoted by $\pi \approx \pi'$, if and only if $T = T'$, and for all $s,t \in T$, $\pi(s) = \pi(t)$ if and only if $\pi'(s) = \pi'(t)$. 
In other words, equivalent partial plans $\pi$ and $\pi'$ 
both assign the same steps $T$ and a set of steps is assigned to a single user by $\pi$ if and only if $\pi'$ also assigns those steps to a single (possibly different) user.

Without loss of generality, 
we may assume the set of steps $S$ is ordered as $s_1,\dots,s_k$.
Each equivalence class $L$ of partial plans corresponds to a unique \emph{pattern} $p$, which is a unique encoding of that class.
More precisely, the \emph{encoding} $p=\mathrm{Enc}(L)$ of an equivalence class $L$ of partial plans for $\approx$ is given by
$p = (T,(x_1,\dots,x_k))$, where for any plan $\pi\in L$, we have that $\pi:T \rightarrow Y$ for some $Y \subseteq U$, and for each $i \in [k]$:
\[
 x_i =
  \begin{cases}
    0 & \text{if $s_i \not\in T$}, \\
    1 & \text{if $i=1$ and $s_1\in T$}, \\
    x_j & \text{if $\pi(s_i) = \pi(s_j)$ and $j < i$}, \\
    \max\set{x_1,\dots,x_{i-1}} + 1 & \text{otherwise}.
  \end{cases}
\]
The vector $(x_1,\dots,x_k)$ is called a \emph{min-vector}, representing a partition of the steps in $T$ such that each block in the partition is assigned to a single user and each block is assigned to a different user.

We will also write $p = \mathrm{Enc}(\pi)$ to represent the fact that $p$ is the pattern of $\pi$, i.e. $p = \mathrm{Enc}(L)$, where $L$ is the equivalence class containing the plan $\pi$.
 A pattern $p$ is {\em authorized} ({\em eligible}, {\em valid}, respectively) if there is a plan $\pi$ which is authorized (eligible, valid, respectively) for which  $p = \mathrm{Enc}(\pi)$.

For our FPT algorithm, we require efficient algorithms for searching and inserting elements into a set of patterns.
Cohen \emph{et al.}~\cite{CoCrGaGuJo13} have shown that such algorithms exist for user-independent constraints, essentially because this set of patterns admits a natural lexicographic ordering.

\subsection {Algorithm for the WSP with user-independent constraints}

The pseudo-code in Algorithm~\ref{alg:pseudo-code} presents an adaptation of the generic FPT algorithm of \cite{CoCrGaGuJo13} to the WSP case with user-independent constraints.
The pseudo-code omits some details concerning heuristics we use to speed up the running time. 
We give details of these heuristics in Section \ref{sec:implementation}.

%

The algorithm considers users one at a time, in some order $u_1, \dots, u_n$. The order of users may be changed dynamically -- that is, after processing user $i$, we may change the ordering of users $u_{i+1}, \dots, u_n$.
We do this in order to move to a possible solution more efficiently by restricting the search space of currently valid patterns.
The details of the dynamic ordering are given in Section \ref{subsec:DynamicOrder}.
As well as prioritising some users by moving them to earlier in the order, we identify some users (called the \emph{useless users}) which we may assume perform no steps at all, and therefore do not need to be processed.
The details about useless users are given in Section \ref{subsec:UU}.

%
%
%
 
 As we process the users, we produce a set $\Pi$ of potential partial solutions and a set $P$ of encodings of these solutions.
More precisely: after processing user $u_i$, the set $\Pi$ is a set of valid partial plans which only assign steps to users from $\{u_1, \dots, u_i\}$. The set $P$ is the set of encodings of plans in $\Pi$, and for each $p \in P$ there is exactly one plan $\pi_p \in \Pi$ such that  $p = \mathrm{Enc}(\pi_p)$. For any valid partial plan $\pi$ that assigns steps only to users from $\{u_1, \dots, u_i\}$, there exists $p \in P$ such that $p = \mathrm{Enc}(\pi)$.
Thus, $P$ is a set of encodings of all valid partial plans with user set contained in $\{u_1, \dots, u_i\}$.
 
At each iteration, having processed users $u_1, \dots u_i$,
we try to assign to the next user $u_{i+1}$ a set $T'$ of steps unassigned by an existing valid plan in $\Pi$, in order to obtain a new valid partial plan. The new user $u_{i+1}$ must be authorized for each step in $T'$.
 At the same time as we construct the new plan, we calculate its pattern. 
 If the resulting plan is eligible (and therefore, valid at this stage), and if its pattern is not already in $P$,
 we add the plan to $\Pi$ and its pattern to $P$. 
 If the resulting plan covers all steps in $S$, we have a solution to the WSP instance.


The overall complexity of the algorithm is determined by $k$, $n$, and the number $w_i$ of patterns (equivalence classes of $\approx$) considered by the algorithm for a pair $(U_i,T)$, where $U_i=\{u_1,\ldots,u_i\}$ is the set of the first $i$ users in iteration, $i=1,\ldots,n$, $T\subseteq S$.
We define the \emph{diversity} of the equivalence relation $\approx$ with respect to an order of users $u_1,\ldots,u_n$ to be $w=\max_{1\le i\le n} w_i$. 
Theorem 1 in \cite{CoCrGaGuJo13} asserts that our algorithm has run-time $O^*(3^k w \log w)$.
Thus, when $w$ is a function of $k$ only, we have an FPT algorithm.
For user-independent constraints, $w \leqslant B_k$ \cite{CoCrGaGuJo13}, where $B_k$ is the $k$th Bell number, and $B_k = 2^{k \log k(1-o(1))}$~\cite{BeTa10}. 

\begin{algorithm}[h!]
\caption{FPT algorithm for the WSP with user-independent counting constraints}\label{alg:pseudo-code}
     \KwIn{Instance of WSP}
     \KwOut{SATISFIABLE and a valid plan, or UNSATISFIABLE.}
\Begin
{

  Initialize the set $\Pi$ of plans with the trivial plan $\pi: \emptyset \rightarrow \emptyset$\;\nllabel{ln:initializePlans}
  Initialize the set $P$ of patterns with the zero-pattern $(\emptyset,(0,\ldots,0))$\;\nllabel{ln:initializePatterns}
  If possible, derive new constraints from given constraints (e.g., see Section \ref{subsec:Pairs})\;\nllabel{ln:markPairs}
  \ForEach{ $u\in U$ \nllabel{ln:user-level} }
  {
    {
      Initialize $\Pi_u=\emptyset$\;\nllabel{ln:big-loop}
      Initialize $P_u = \emptyset$\;
      Preprocess constraints  (see Section \ref{subsec:eligCheck})\nllabel{ln:outerPreprocess}\;
      \ForEach{ pattern $p = (T,(x_1,\dots,x_k))$ in $P$ }
      {
	$T_u \leftarrow A(u)\setminus T$\; \nllabel{ln:is-plan-auth}
	  \ForEach{ $\emptyset \neq T' \subseteq T_u$ }
	  {
	    Let $\pi_p$ be a plan with pattern $p$ in $\Pi$\;
    	    $\pi' \leftarrow \pi_p \cup (T' \rightarrow u)$\;
	    Preprocess constraints (see Section \ref{subsec:eligCheck})\nllabel{ln:innerPreprocess}\;  
	    Check whether $\pi'$ is eligible and propagate constraints  (see Sections \ref{subsec:CP} and \ref{subsec:Pairs})\nllabel{ln:propagateConstraints}\;
	    \If{$\pi'$ is eligible \nllabel{ln:is-plan-valid}}
	    {
	      \eIf{ $T \cup T' = S$ }
	      {
		\Return SATISFIABLE and $\pi'$\;
	      }
	      {
		Compute the pattern $p'$ for $\pi'$\nllabel{ln:compute-pattern}\;
		\If{$p' \not\in P \cup P_u$\nllabel{ln:add-or-not}}
		{
		  Add $\pi'$ to $\Pi_u$\;
		  Add $p'$ to $P_u$\;
		}
	      }
	    }
	  }
      }
    }
  \eIf{ $\Pi_u = \emptyset$\nllabel{ln:nothing} }
  {
   \ForEach{$u'$ such that $A(u') \subseteq A(u)$}
   {
      $U \leftarrow U \setminus \{u'\}$ \text{(Remove useless users, see Section \ref{subsec:UU})}\;\nllabel{ln:useless}
   }
  }
  {
   $\Pi \leftarrow \Pi \cup \Pi_u$\nllabel{ln:extension}\;
   $P \leftarrow P \cup P_u$\;
  }
  Choose a user for the next iteration (see Section \ref{subsec:DynamicOrder})\;\nllabel{ln:re-order-users}
  }
 \Return UNSATISFIABLE;
}
\end{algorithm}

\section{Implementing the FPT algorithm}\label{sec:implementation}

We provide more details of our implementation of Algorithm 1 below. 
For our experiments, we use  not-equals, at-most-$r$, and at-least-$r$ constraints. A \emph{not-equals constraint} $(s,t,\neq)$ is specified by a pair of steps $s$ and $t$; a plan $\pi$ satisfies the constraint $(s,t,\neq)$ if $\pi(s) \neq \pi(t).$ An {\em at-most-$r$ constraint} may be represented as a tuple $(r,Q,\leqslant)$, where $Q \subseteq S$ and $1\leqslant r \leqslant |Q|$, and is satisfied by any plan that allocates no more than $r$ users in total to the steps in $Q$.
An {\em at-least-$r$ constraint} may be represented as $(r,Q,\geqslant)$ and is satisfied by any plan that allocates at least $r$ users to the steps in $Q$.
Note that these at-most and at-least constraints impose ``confidentiality" and ``diversity" requirements on the workflow, which can be important in a business environment.

Our implementation therefore includes some heuristics specifically designed for these constraints.
In this section we describe the main ideas used for implementing the FPT algorithm, and heuristic speed-ups that have been introduced to make implementations competitive with and more efficient than SAT4J on the reduction to the PB SAT problem of Theorem~\ref{prop:red2}.
Note that only the heuristics referred to in lines \ref{ln:outerPreprocess}, \ref{ln:innerPreprocess} and \ref{ln:propagateConstraints} are specific to  not-equals, at-most-$r$, and at-least-$r$ constraints. Without these heuristics, Algorithm 1 becomes a generic algorithm for any user-independent constraints.

\subsection{Preprocessing and eligibility checking} \label{subsec:eligCheck}
Not-equals and at-least-$r$ constraints are preprocessed in the outer loop on a per-user basis in line~\ref{ln:outerPreprocess}: only at-least-$r$ constraints containing some steps authorized to the current user and not-equals constraints with both steps authorized to the current user are passed into the inner loops for eligibility checking and propagation. At-least-$r$ constraints are also preprocessed per-pattern to guarantee their efficient checking in the inner loop (line~\ref{ln:innerPreprocess}): only at-least-$r$ constraints containing steps from $T'$ are chosen to be used for eligibility checking.
At-most-$r$ constraints are not preprocessed, but checked directly for violation and propagated in combination with not-equals constraints as explained later (line~\ref{ln:propagateConstraints}). 

Much of the work of the implementation of the algorithm is done in line~\ref{ln:is-plan-valid}: we consider a plan $\pi_p$ with a given valid pattern $p$ and test whether its authorized extension (by the assignment of steps in $T'$ to user $u$) is eligible. Line~\ref{ln:is-plan-auth} guarantees that we deal only with authorized plans. 
In computational experiments, 
at-most-$r$ and not-equals constraints reject a much larger proportion of candidates for a valid plan than at-least-$r$ constraints. In other words, they are much easier to violate than at-least-$r$ constraints. Since not-equals constraints are preprocessed per-user for efficient checking, and at-most-$r$ constraints are not preprocessed but used for propagation and more efficient dynamic iteration by users,  the constraints are checked in the following order:
\begin{inparaenum}[(1)]
 \item not-equals constraints;
 \item at-most-$r$ constraints;
 \item at-least-$r$ constraints.
\end{inparaenum}
At the same time as we check at-most-$r$ constraints, we also propagate at-most-$r$  constraints and pairs of intersecting at-most-$r$ constraints; see the next two sections.
The pattern for the extended valid plan is computed and added to the set $P_u$ of extended patterns in line~\ref{ln:compute-pattern}.
Results by Cohen et al.~\cite{CoCrGaGuJo13} assert that these subroutines can be performed efficiently.

\subsection{Constraint propagation}\label{subsec:CP}
After checking that a partial plan $\pi \cup (T' \rightarrow u)$ does not violate at-most-$r$ constraints directly, we propagate information (line~\ref{ln:propagateConstraints}) about the current state of any at-most-$r$ constraint
as follows.

Suppose $\ell$ steps in a constraint $c=(r,Q,\leqslant)$ have been assigned to $r-1$ distinct users in $\pi \cup (T' \rightarrow u)$, and $\ell \le |c|-2$.
Then the remaining $q=|c| - \ell\ge 2$ steps in $Q$ must be assigned to a single user.
If there is no remaining user $u''$ authorized for all $q$ unassigned steps in $Q$, we discard the partial plan $\pi \cup (T' \rightarrow u)$ and its pattern.

Similarly, if any pair of the $q$ unassigned steps in $Q$ is the scope of some not-equals constraint, we discard the partial plan and its pattern. 
On the other hand, if such a user $u''$ authorized for all the unassigned steps in an at-most-$r$ constraint $c$ is found, it is called \emph{useful} and may be given priority to be used in the next iteration.
Pairs of intersecting at-most-$r$ constraints are also propagated 
at this time; see the next section.

\subsection{Propagating pairs of intersecting at-most-$r$ constraints}\label{subsec:Pairs}
The latest implementations of the algorithm propagate information about pairs of intersecting at-most-$r$ constraints -- that is, pairs of constraints with overlapping sets of steps (earlier implementations considered only pairs of at-most-$r$ constraints intersecting in at least two steps). 
In line~\ref{ln:markPairs}, before starting the iteration, pairs of intersecting at-most-$r$ constraints are recorded separately, with one of the common steps marked.
The pairs are propagated
during line \ref{ln:propagateConstraints}
when the marked common step is unassigned. In this case, we need to find at least one user $u''$ authorized for all unassigned steps in both constraints compising the pair as follows.

Suppose $c_1=(r_1,Q_1,\leqslant)$ and $c_2=(r_2,Q_2,\leqslant)$ are the two constraints comprising a pair, $Q_1\cap Q_2\not= \emptyset$, and the marked intersection step is $s\in Q_1\cap Q_2$.
As above, suppose $\ell_i$ steps in a constraint $c_i=(r_i,Q_i,\leqslant)$ are assigned to $r_i-1$ distinct users in $\pi \cup (T' \rightarrow u)$, $\ell_i \le |c_i|-2$, and
the remaining $q_i=|c_i| - \ell_i \ge 2$ steps in $Q_i$ must be assigned to a single user, $i=1,2$. Moreover, suppose the step $s\in Q_1\cap Q_2$ is unassigned in $\pi \cup (T' \rightarrow u)$. Then the $q_1$ unassigned steps in $c_1$ and the $q_2$ unassigned steps in $c_2$ must be assigned to the same single user $u''$ in a later iteration. Furthermore, $u''$ must be authorized for all the unassigned steps in $c_1$ and $c_2$. If such a user $u''$ is not available among the remaining users, or if any pair of the unassigned steps in $Q_1\cup Q_2$ is the scope of some not-equals constraint, we discard the partial plan $\pi \cup (T' \rightarrow u)$ and its pattern. 
On the other hand, a user $u''$ authorized for all the unassigned steps in a pair of intersecting at-most-$r$ constraints is called \emph{super-useful} and may be given priority to be used for the next iteration.


\subsection{Useless users}\label{subsec:UU}
Each iteration of the algorithm considers assigning some steps to a particular user $u$ and constructs a set $P_u$ of extended valid patterns and, respectively, a set $\Pi_u$ of partial plans that include this user $u$. The construction of $P_u$ and $\Pi_u$ is based on patterns in the set $P$ generated after the previous iterations, constraints in $C$, and the list of authorizations $A(u)$ of $u$.
Suppose $\Pi_{u'}$ is a set of plans in Algorithm~\ref{alg:pseudo-code} used for a later iteration, when another user $u'$ is considered, and $A(u')\subseteq A(u)$. Then, $\Pi_u=\emptyset$ implies $\Pi_{u'}=\emptyset$, i.e. such a user $u'$ can be disregarded later in the iteration.  
The reason for this is that any steps assigned to $u'$ could instead be assigned to $u$.
We justify this claim formally with the following result.

\begin{thm}\label{prop:useless}
Suppose Algorithm~\ref{alg:pseudo-code} considers users in the order $u_1,u_2,\ldots,u_n$ to solve a WSP instance $(S,U,\mathcal{A},C)$ with user-independent constraints $C$.
Suppose also that
$\Pi_{u_i}=\emptyset$ for some user $u_i$ (line~\ref{ln:nothing}), $i\ge 1$,
and a user $u_j\in \{u_{i+1},\ldots,u_n\}$ has $A(u_j)\subseteq A(u_i)$. 
Then  
 $\Pi_{u_j}= \emptyset$.
\end{thm}

\pf
Suppose for a contradiction that $\Pi_{u_j}\neq \emptyset$.
Then $\Pi_{u_j}$ contains a valid partial plan $\pi: T \rightarrow \{u_1, \dots, u_j\}$ for some $T \subseteq S$. 
By definition of $\Pi_{u_j}$,
$\pi^{-1}(u_j) \neq \emptyset$, and $\pi \not\approx \pi'$ for any valid partial plan $\pi': T \rightarrow \{u_1, \dots, u_{j-1}\}$
(as otherwise $\pi$ would not have been added to $\Pi_{u_j}$).


Let $\pi_{i}$ be the plan $\pi$ restricted to $\{u_1,\dots, u_i\}$, i.e. $\pi_{i} : T' \rightarrow \{u_1, \dots, u_i\}$, 
where $T'=\pi^{-1}(\{u_1, \dots , u_i\})$.
Clearly, $\pi_{i}$ is a valid partial plan that corresponds to a pattern $p$ encoding the equivalence class $L$ of $\pi_{i}$. Since $\Pi_{u_i}=\emptyset $ in line~\ref{ln:nothing} of Algorithm~\ref{alg:pseudo-code}, there is an authorized and eligible partial plan $\pi_{i-1}\in \Pi$ having the same pattern $p$ as $\pi_i$ that does not use the user $u_i$, i.e. $\pi_{i-1} : T' \rightarrow \{u_1, \dots, u_{i-1}\}$, and $\pi_{i-1}\approx\pi_{i}$.

Now let $\pi_{i+1}$ be  $\pi$ restricted to $\{u_{i+1},\dots, u_j\}$, i.e. $\pi_{i+1} : T\backslash T' \rightarrow \{u_{i+1},\dots, u_j\}$, where the set of steps $T'$ is the same as covered by $\pi_i$ and $\pi_{i-1}$. Consider $\pi'=\pi_{i-1}\cup\pi_{i+1}$. Since $\pi=\pi_i\cup\pi_{i+1}$ 
and
$\pi_{i-1}\approx\pi_{i}$,
we have that $\pi \approx \pi'$.
It follows that $\pi'$ is eligible and, by construction, $\pi'$ is authorized. 
Therefore we have a valid partial plan $\pi': T \rightarrow \{u_1, \dots, u_j\}$ such that $\pi' \approx \pi$ and ${\pi'}^{-1}(u_i)=\emptyset$.
 
Finally, let $\pi''$ be a plan obtained from $\pi'$ by reassigning all the steps assigned to $u_j$ in $\pi'$ to the user $u_i$. Clearly, this is possible because $A(u_j)\subseteq A(u_i)$. In other words, we respect authorizations in $\pi''$ and have $\pi''^{-1}(u_i)=\pi'^{-1}(u_j)$, $\pi''^{-1}(u_j)=\emptyset$. Since $\pi''$ can be obtained from $\pi'$ by a permutation of two users, $u_i$ and $u_j$, and all the constraints are user-independent, $\pi''$ is eligible
and $\pi'' \approx \pi' \approx \pi$.

Thus, we have a valid partial plan $\pi'':T \rightarrow \{u_1, \dots, u_{j-1}\}$ such that $\pi''\approx \pi$, a contradiction.
\qed \\

We say a user $u'$ is {\em useless} for the current iteration if there exists a user $u$ such that $A(u') \subseteq A(u),$ $u$ has been considered in iteration earlier, and $\Pi_u = \emptyset$ in line~\ref{ln:nothing} of Algorithm~\ref{alg:pseudo-code}.
Theorem~\ref{prop:useless} implies that, if we discover a useless user $u'$ in Algorithm~\ref{alg:pseudo-code}, without loss of generality, we may assume that there are no steps assigned to this user $u'$ in a final solution (a complete valid plan).
One of the heuristic speed-ups we employ is to identify and ignore useless users (line~\ref{ln:useless}).
This is the only heuristic speed-up used in implementations of Algorithm~\ref{alg:pseudo-code} that has a deterministic nature and requires a proof of its correctness.

\subsection{Dynamic choice of users in iteration (useful users)}\label{subsec:DynamicOrder}
In an effort to satisfy some of the at-most-$r$ constraints and to reduce the number of unassigned steps in partial plans as quickly as possible, useful or super-useful users identified during the propagation in line  \ref{ln:propagateConstraints} are used to perform the next iteration. The priority can be given, for example, to a useful user satisfying at least two at-most-$r$ constraints independently, or the first detected super-useful user, or a super-useful user covering the largest number of steps in the corresponding at-most-$r$ constraints, etc. 
The chosen useful user is moved to the beginning of the list of remaining users, and the list of remaining users is adjusted accordingly (line~\ref{ln:re-order-users}). 
This determines a dynamic ordering on the set of users through which the algorithm iterates.
Depending on the choice of useful users, implementations of the FPT algorithm can behave very differently and competitively with respect to each other, making it difficult to select the best implementation among several possible choices.
The implementation chosen for the experiments in Section \ref{sec:experiments} uses the first super-useful user detected during propagation, if such a user exists, and otherwise uses the first useful user detected.

\section{The WSP as a pseudo-Boolean SAT problem}\label{sec:solving-wsp:pb-boolean}

In this section we describe how the WSP with user-independent counting constraints can be encoded as a pseudo-Boolean SAT problem and prove correctness of the encoding.
Wang and Li \cite{WaLi10} encoded their WSP instances as a pseudo-Boolean SAT problem and used a PB SAT solver (SAT4J) to solve them.
Pseudo-Boolean SAT solvers are recognized as an efficient way to solve general constraint networks \cite{BePa10}.
In their experiments,  Wang and Li \cite{WaLi10} considered not-equals constraints. They also considered a number of other constraints, 
which we do not use in our experimental work since they add little complexity to a WSP instance. 
In the experiments of Wang and Li, SAT4J solved all generated instances quite efficiently. 

We test SAT4J on a set of WSP instances of a different type, where SAT4J's effectiveness and efficiency vary a lot. We show that our reduction to the PB SAT problem and SAT4J can be still successfully applied to lightly-constrained instances, which are not likely to be unsatisfiable. We use not-equals constraints as well as  at-most-$r$ and at-least-$r$ constraints.
For convenience and by an abuse of notation, given a constraint $c$ of the form $(r,Q, \geqslant)$ or $(r,Q, \leqslant)$, we will write $s \in c$ to denote that $s \in Q$, and define $|c|$ to be $|Q|$. 

An authorization list for a step $s$ is a set of users authorized to perform $s$ and is denoted by $A(s)=\{ u \in U:\ s \in A(u)\}$.
We
encode such constraints in the same way as
Wang and Li \cite{WaLi10}, by defining a binary variable $x_{u,s}$ for every pair $(u,s)$ such that $u$ is authorized for $s$. That is, a variable $x_{u,s}$ is defined if and only if $s \in A(u)$.
In addition, for each at-least-$r$ constraint $c$ and user $u$, we introduce a
(0,1)-variable $z_{u,c}$, and, for each at-most-$r$ constraint $c$ and user $u$, we
introduce a (0,1)-variable $y_{u,c}$. 
These variables are subject to the following constraints:
   \begin{enumerate} 
   \item[(PB1)] for each step $s$:\ \ \ $\sum_{u \in A(s)} x_{u,s} = 1$;
   \item[(PB2)]  for each not-equals constraint $(s,t,\ne)$ and user $u\in A(s)\cap A(t)$:\ \ \ 
\mbox{$x_{u,s} + x_{u,t} \leqslant 1$};
   \item[(PB3)]  for each at-least-$r$ constraint $c$ and user $u$:\ \ \ 
   $z_{u,c} \leqslant \sum_{s
\in A(u) \cap c}x_{u,s}$;
   \item[(PB4)]  for each at-least-$r$ constraint $c$:\ \ \ 
   $\sum_{u \in U} z_{u,c} \geqslant r$;
   \item[(PB5)]  for each at-most-$r$ constraint $c$,  $s \in c$, and $u \in A(s)$:\ \ \ 
   $x_{u,s} \leqslant y_{u,c}$;
   \item[(PB6)]  for each at-most-$r$ constraint $c$:\ \ \ 
   $\sum_{u \in U} y_{u,c} \leqslant r$.
  \end{enumerate}
 
The goal of a PB SAT solver is to find an assignment of values to these variables, representing a plan, where $x_{u,s} = 1$ if and only if user $u$ is assigned to step $s$.
Informally, (PB1) ensures that each step is assigned to a single user; (PB2) ensures that all not-equals constraints are satisfied; (PB3) and (PB4) ensure that all at-least-$r$ constraints are satisfied; and (PB5) and (PB6) ensure that all at-most-$r$ constraints are satisfied.

\begin{thm}\label{prop:red2}
A WSP instance with not-equals, at-most-$r$ and at-least-$r$ constraints has a solution if and only if the corresponding pseudo-Boolean SAT problem (PB1)-(PB6) has a solution.
\end{thm}

\pf
Suppose first that our WSP instance with not-equals, at-most-$r$, and at-least-$r$ constraints has a solution. 
Then set $x_{u,s} = 1$ if $u$ is assigned to $s$ in the WSP solution (noting that
$x_{u,s}$ is defined, as $u$ must be authorized for $s$), and set $x_{u,s}=0$
otherwise.
For each at-least-$r$ conststraint $c$, let $z_{u,c}=1$ if and only if $u$ is assigned
to a step in $c$.
Similarly, for each at-most-$r$ constraint $c$, let $y_{u,c} = 1$ if and only if $u$ is
assigned to a step in $c$. 
It is easy to see that this assignment of $(0,1)$-values to variables $x_{u,s}, z_{u,c}$, and $y_{u,c}$ satisfies all of the pseudo-Boolean constraints in (PB1)-(PB6).
%

Conversely, suppose our pseudo-Boolean SAT problem (PB1)-(PB6) has a solution. 
By the first set of PB constraints, for each step $s$ there exists a unique user $u$
such that $x_{u,s}=1$. 
Consider the solution to the WSP in which each step $s$ is assigned to the unique $u$
such that $x_{u,s}=1$. 
As $x_{u,s}$ is only defined for authorized pairs $(u,s)$, this solution only
assigns users to steps for which they are authorized. 
We now show that this solution satisfies all the constraints in the WSP instance.

For an inequality constraint $(s,t,\ne)$, we have that, for each user $u$, either $u$
is not authorized for at least one of $s$ and $t$, or $x_{u,s} + x_{u,t} \leqslant
1$. 
It follows that no user performs more than one of $s$ and $t$, so the corresponding
not-equals constraint is satisfied.

For an at-least-$r$ constraint $c$, the satisfied inequality in (PB4) guarantees that at least $r$ of variables $z_{u,c}$, $u\in U$, are set equal to $1$.
Observe that $z_{u,c}=1$ implies that
$u$ performs a step in $c$: since $z_{u,c} \leqslant \sum_{s \in A(u)\cap c}x_{u,s}$ in (PB3), it must be the case that $x_{u,s}=1$ for some $s \in A(u)\cap c$.
Thus, as we have $\sum_{u \in U} z_{u,c} \geqslant r$, it follows that there are at
least $r$ users that perform a step in $c$, and so $c$ is satisfied.

For each at-most-$r$ constraint $c$, the equalities in (PB1) imply that
for each $s \in c$, there exists $u \in A(s)$ such that $x_{u,s}=1$,
i.e. $s$ is performed by a certain authorized user $u$.
Since the inequalities in (PB5) are satisfied,
$x_{u,s} \leqslant y_{u,c}$ implies $y_{u,c}=1$ for that particular user $u$.
Thus, $y_{u,c}=1$ for every user $u$ that performs at least one step in $c$.
Since the inequality in (PB6) is satisfied, at most $r$ users perform a step in $c$, and
so $c$ is satisfied.
\qed


\section{Experiments}\label{sec:experiments}

Due to the difficulty of acquiring real-world workflow instances, Wang and Li \cite{WaLi10} used synthetic data in their experimental study. We follow a similar approach to test experimentally the FPT algorithm and the reduction to the PB SAT problem.
We use C++ to implement the FPT algorithm and to encode the reduction for the WSP with user-independent counting constraints.%
\footnote{
We would like to emphasize that even though the constraints considered in the theoretical part of Wang and Li's paper are purely user-independent, the authors consider randomly generated relations between users for their experiments. Therefore the experimental tests in \cite{WaLi10} are done not in a user-independent environment.
}
We generate a number of random WSP instances with not-equals and counting constraints and compare the performance of one of our implementations of the FPT algorithm with that of SAT4J on the reduction when solving the same instances.
All our experiments use a MacBook Pro computer having a 2.6 GHz Intel Core i5 processor, 8 GB 1600 MHz DDR3 RAM\footnote{Our computer is more powerful than the one used by Wang and Li \cite{WaLi10}.} and running Mac OS X 10.9.5. More experimental test results of earlier versions and implementations of this FPT algorithm and the reduction to the PB SAT problem using SAT4J can be found in \cite{FAW2014}. 

\subsection{Testbed Design}	

An authorization list for a step $s\in S$ is a set of users $A(s)\in U$, authorized to perform $s$.
The set of authorization lists $\set{A(s) : s\in S}$ can be thought of as $\set{A(u) : u\in U}$, where $A(u) = \set{s\in S : u \in A(s)}$. 
We assumed that every user was authorized for at least one step but no more than $\lceil\frac{k}{2}\rceil$ steps; that is, $1 \leqslant |A(u)| \leqslant \lceil\frac{k}{2}\rceil$.

All constraints are of the form not-equals, at-most-$r$, or at-least $r$, where $r$ is some small number. The not-equals constraints have domain of size $2$, while the at-most-$r$ and at-least $r$ constraints have domain of size $t$, where $t$ is some number bigger than $r$.

We vary the number of not-equals constraints as a percentage $d$ of $\binom{k}{2}$, the maximum possible number of these constraints. 
For these experiments, we use the same number of at-most-$r$ constraints as the number of at-least-$r$ constraints, denoted by $b$.
All not-equals, at-most-$r$, at-least-$r$ constraints and authorizations are generated for each instance separately, uniformly at random.%
\footnote{This experimental setup is different from the one used in our earlier work~\cite{FAW2014}.}


Counting at-most-$r$ and at-least-$r$ constraints are generated by first enumerating all possible $t$-element subsets of $S$ using an algorithm from Reingold {\em et al.}~\cite{ReNiDe77}.
Then Durstenfeld's version of the Fisher-Yates random shuffle algorithm~\cite{Du64,FiYa49} is used to select separately and independently at random $t$-element subsets of $S$ for the scopes of at-most-$r$ and at-least-$r$ constraints, $b$ subsets for each.
The random shuffle algorithm is also used to select steps in $A(u)$ for which each user $u$ is authorized, and the list of authorization sets $\set{A(u) : u\in U}$ is generated uniformly at random, subject to the cardinality constraints $1 \leqslant |A(u)| \leqslant \lceil\frac{k}{2}\rceil$.
Finally, the random shuffle is used to randomly select not-equals constraints, taking $d$\% of the set of all possible not-equals constraints.

\subsection{Testbed Choice}	

For the experiments, we wanted to use WSP instances that were both simple and challenging to solve, and close to what might be actually expected in practice.
Therefore, we restricted our attention to counting constraints with $r=3$ and $t=5$, calling them \emph{at-most-3} and \emph{at-least-3 constraints}, respectively. The rationale behind this is as follows.
From our ad hoc experiments, at-most-$r$ and at-least-$r$ constraints with $r=1$ or $2$ make the problem instances easily solvable by both SAT4J and the FPT algorithm. On the other hand, at-most-$r$ or at-least-$r$ constraints with $r\geqslant 4$ are not likely to appear in practice. Similarly, at-most-$r$ and at-least-$r$ constraints with $t\leqslant 4$ appeared to provide relatively easy solvable WSP instances, and $t\geqslant 6$ seems to be less likely to appear in practice. Therefore, for simplicity and to keep the things challenging and realistic enough at the same time, the choice of $r=3$ and $|c|=5$ seems to be well justified. Increasing $r$ and $t$ and keeping them close to each other seem to provide more challenging instances, which are left out of the scope of this paper. On the other hand, smaller values of $r$ and $t$, or a larger difference between them, seem to create WSP instances instances whose satisfiability is easier to decide.

Based on what might be expected in practice, we used values of $k=15, 20, 25$ for the number of steps, and set $n = 10k$ for the number of users.\footnote{Schaad {\em et al.} investigated several case studies in which authorization constraints were relevant, including a loan origination process in a bank~\cite{ScLoSo06} and the creation of electronic signatures in a law practice~\cite{ScSpWe05}.  These two business processes used 13 and 12 steps, respectively.}
For the percentage $d$ of not-equals constraints (out of the $\binom{k}{2}$ possible not-equals constraints), we used values of $d=10,20,30$\,(\%).

For convenience, we adopt the following convention to label our test instances: $b.d$ denotes an instance with $b$ at-most-3 constraints, $b$ at-least-3 constraints, and not-equals constraint density $d$ (e.g., see Tables~\ref{tbl:results-k-equal-20} and \ref{tbl:results-k-equal-25}).

Considering $b$ and $d$ as parameters, we try to explore an area where both satisfiable (sat) and unsatisfiable (unsat) instances of the WSP are relatively likely to occur.
Starting with instances having not too many not-equals constraints, informally, we expect that the difficulty of solving instances $b.d$ for fixed $k$ and $b$ would increase as $d$ increases (as the problem becomes ``more constrained'').
Similarly, for fixed $k$ and $d$, we expect instances $b.d$ would become harder to solve as $b$ increases.
We also expect that the time taken to solve an instance would depend on whether the instance is satisfiable or not, with unsatisfiable instances requiring a solver to examine all possible plans (or to provide a certificate of their unsatisfiability).

We generate a set of instances of different degrees of hardness by varying the not-equals constraint density and the number of counting constraints. 
The resulting set of test instances includes some that are satisfiable with a relatively high diversity $w$ of their search space (``lightly-constrained''), and some that are either satisfiable with a relatively low diversity $w$ or are unsatisfiable (``well-constrained'').
For each number of steps, $k=15,20,25$, and three different constraint densities, $d=10,20,30$\,(\%), to determine the range of values for $b$ (which is dependent on $k$), we start with instances that seem to be lightly-constrained and normally can be efficiently solved by SAT4J.
We then gradually increment $b$ in steps by $2$ and generate more instances.
We stop generating instances when we find two consecutive values of $b$ for which no instances are satisfiable.
The minimum and maximum values of $b$ used in the experiments to generate Tables~\ref{tbl:results-different-k}--\ref{tbl:results-k-equal-25} correspond to instances that we view as borderline.
In other words, we start with instances that are likely to be lightly-constrained and stop at instances which are likely to be well-constrained as the corresponding three instances for the same values of $k$ and $b$ are unsatisfiable.
For the test instances in this paper, $b$ ranges from $2$ to $32$ for $k=15$, from $10$ to $38$ for $k=20$, and from $22$ to $36$ for $k=25$, respectively.
The interested reader is referred to \cite{FAW2014} for further details about possible selection of parameters for test instances.

\subsection{Results}

In the experiments we compare the run-times and performance of an implementation of our FPT algorithm, called \emph{Solver FPT}, and of SAT4J on the reduction to the PB SAT problem (described in Section \ref{sec:solving-wsp:pb-boolean}), referred to as \emph{Solver SAT4J}. Overall, Solver FPT was able to solve all the $117$ test instances, while Solver SAT4J solved only $100$ of them ($85.5\%$). See Table~\ref{tbl:results-different-k} for overall statistics: the corresponding numbers of unsolved instances are in parenthesis. For average time values, we assume that the running time on the unsolved instances can be considered as a lower bound on the time required to solve them. Therefore average time values in Table~\ref{tbl:results-different-k} take into consideration unsolved instances for Solver SAT4J: they are estimated lower bounds on its average time performance.

\begin{table}[h]\centering \small\setlength{\tabcolsep}{3pt}\setlength{\extrarowheight}{.1pt}
\caption{Summary statistics for $k \in \set{15,20,25}$}\label{tbl:results-different-k}
\begin{tabular}{|r|r|r|r|r|r|r|r|r|}
\cline{4-7}
  \multicolumn{3}{c|}{} & \multicolumn{2}{c|}{\bf Solver SAT4J} & \multicolumn{2}{c|}{\bf Solver FPT} \\
\hline
{\footnotesize \#Steps} & {\footnotesize Interval for $b$} & {\footnotesize Instance Type} & {\footnotesize \#Instances} & {\footnotesize Mean Time (s)} & {\footnotesize \#Instances} & {\footnotesize Mean Time (s)} \\
\hline
15 & $2\le b\le 32$ & sat & 26 & 1.25 & 26 & 0.74  \\
     &	& unsat & 22 & 327.84 & 22 & 0.32 \\
     \cline{3-7}
     &	& all & 48 & 150.94 & 48 & 0.55\\
\hline
\hline
20 & $10\le b\le 38$ & sat & 18 & 38.04 & 18 & 20.77\\
      & & unsat & 21\,(6) & 1,096.21 & 27 & 20.43\\
     \cline{3-7}
     &	& all & 39\,(6) & 672.94 & 45 & 20.56 \\      
\hline
\hline
25 & $22\le b\le 36$ & sat & 5\,(2) & 913.02 & 7  & 3,173.99\\
      & & unsat & 8\,(9) & 1,724.44 & 17 & 845.41\\
     \cline{3-7}
     &	& all & 13\,(11) & 1,487.78 & 24 & 1,524.58 \\      
\hline
\end{tabular}
\end{table}



For the number $k$ of steps equal to $15$, in general, although SAT4J was able to solve all the instances, Solver FPT was several hundred times more efficient than SAT4J, with a much lower 
standard deviation
in time performance 
($0.98$ sec versus $ 293.1$ sec, respectively).
However, for lightly-constrained instances ($b=2$ and $4$), Solver SAT4J was usually about one order of magnitude more efficient than Solver FPT. Nevertheless, both solvers solved the WSP instances in seconds or tenth of seconds, and so Solver FPT could be successfully used for these kind of lightly-constrained instances as well.
Overall, average performance and time variance of both solvers on satisfiable instances was similar, with Solver FPT having a small advantage.
For unsatisfiable instances, Solver FPT appeared to be about three orders of magnitude more efficient than SAT4J. Since we do not know in advance whether a given instance is satisfiable or not, Solver FPT can be considered as clearly superior, and should be applied to instances with $b\ge 6$, while Solver SAT4J seems to be a better choice when $b\le 5$. Also, for small values of $b$ ($b\le 10$), SAT4J seems to be slightly more efficient for lower densities of not-equals constraints, $d\le 10\%$.

\medskip

\begin{figure}[h]
	\centerline {\includegraphics[width=5.7in]{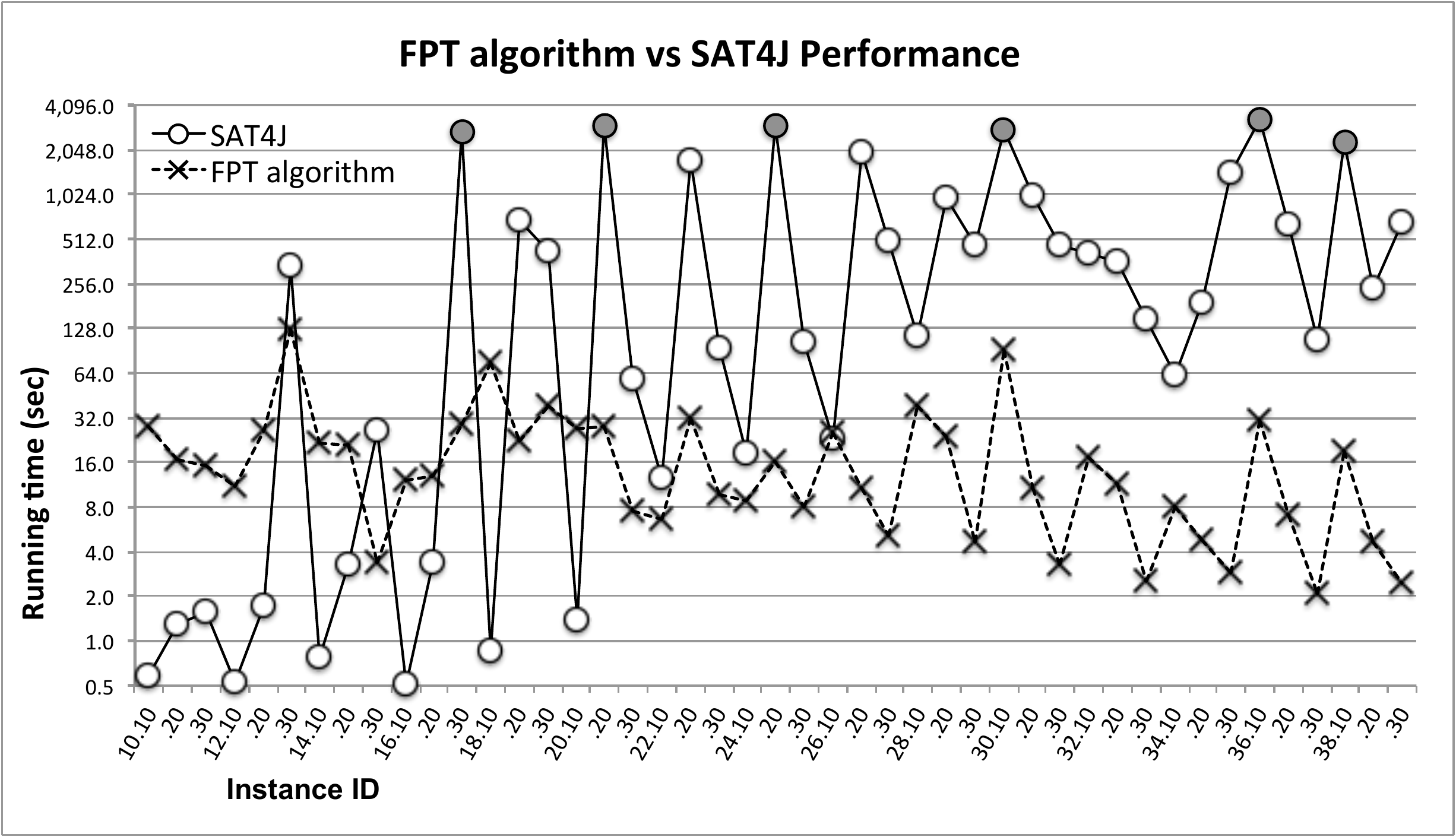}}
	\caption{Test results for $k=20$ steps\label{fig:20}}
\end{figure}

\begin{figure}[h]
	\centerline {\includegraphics[width=4.8in]{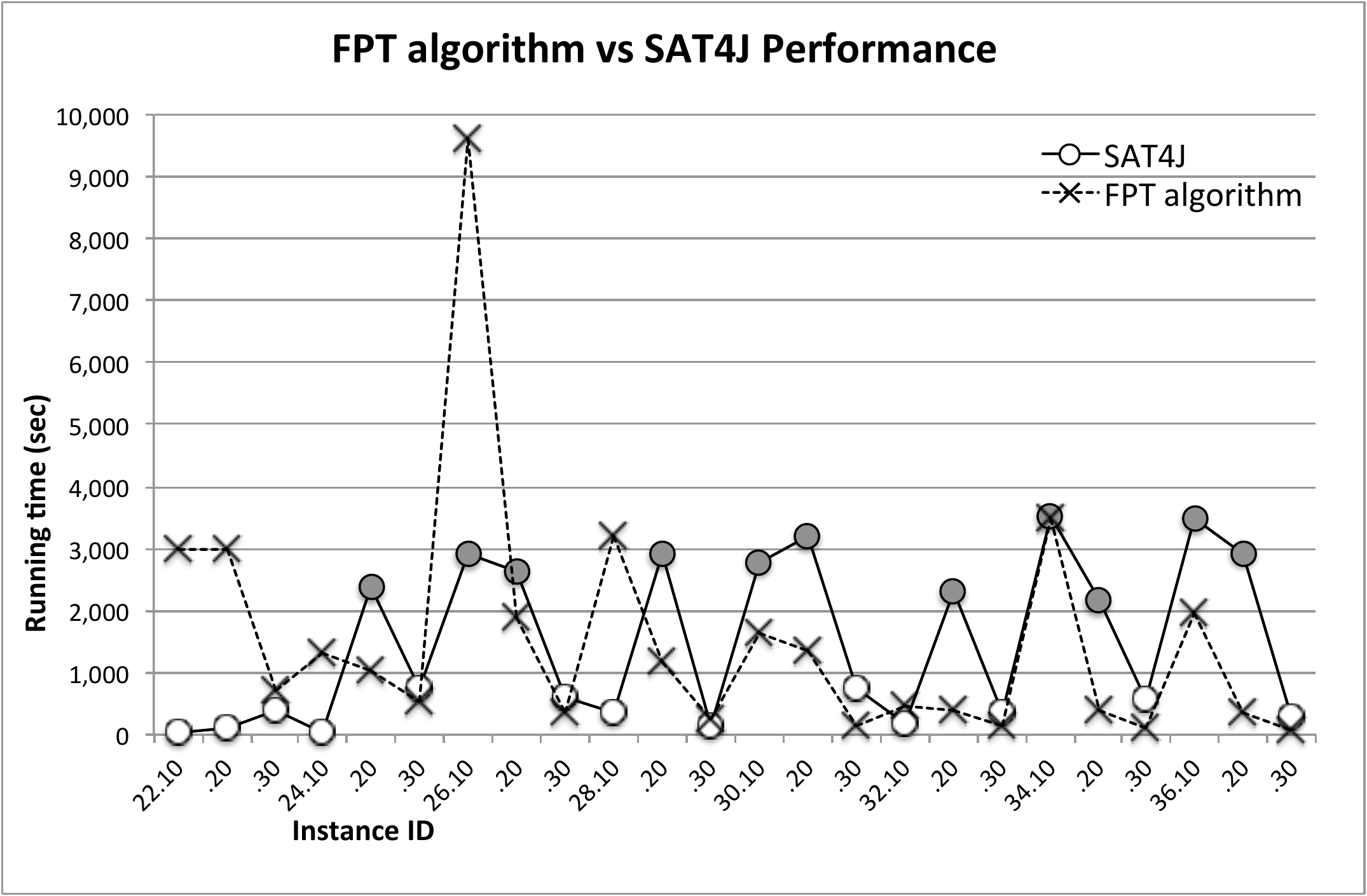}}
	\caption{Test results for $k=25$ steps\label{fig:25}}
\end{figure}

For the number $k$ of steps equal to $20$ and $25$, we provide Tables \ref{tbl:results-k-equal-20} and \ref{tbl:results-k-equal-25}, respectively, which give detailed results of our experiments, and
Figures~\ref{fig:20} and \ref{fig:25}, respectively, depicting the results graphically. 
Notice that Solver FPT reaches a conclusive decision (sat or unsat) in all cases, whereas SAT4J fails to reach such a decision for some instances, typically because the machine runs out of memory.
In Figures~\ref{fig:20} and ~\ref{fig:25}, the shaded circles represent the instances unsolved by SAT4J.
In Tables~\ref{tbl:results-k-equal-20} and \ref{tbl:results-k-equal-25}, we record the outcomes and CPU time taken by the two solvers to run on each instance. The best running time for each instance is in bold.
Auxiliary information for Solver FPT includes the number of users considered and the number of patterns in the search space (a measure of the diversity $w$ of the instance) generated before a valid plan was obtained or the instance was recognized as unsatisfiable.
Also, for unsat instances, we show the number of users $n_w$ that could not extend the set of patterns in line~\ref{ln:extension} of Algorithm~\ref{alg:pseudo-code} (i.e., for such users $u$, $P_u=\emptyset$ in line~\ref{ln:nothing}) and, respectively, the number of useless users $n_u$, whose authorization lists are dominated by those $n_w$ users. We use notation $n: n_w\rightarrow n_u$, where $n$ is 
the total number of users.

\begin{table}[!th]\centering\small\setlength{\tabcolsep}{3pt}
\caption{Experimental test results for $k = 20$} \label{tbl:results-k-equal-20}
\begin{tabular}{|r|c|r||c|r|r|r|}
\cline{2-7}
  \multicolumn{1}{c}{} & \multicolumn{2}{|c||}{\bf Solver SAT4J} & \multicolumn{4}{c|}{\bf Solver FPT} \\
\hline
{\footnotesize Instance ID} &{\footnotesize Output} &{\footnotesize CPU Time\,(s)} &{\footnotesize Output} &{\footnotesize CPU Time\,(s)} &{\footnotesize \#Users} &{\footnotesize \#Patterns} \\
\hline
10.10&sat&{\bf 0.58}&sat&28.18&6&2,286,676	\\
10.20&sat&{\bf 1.29}&sat&16.46&7&497,634	\\
10.30&sat&{\bf 1.55}&sat&15.07&17&177,426	\\
\hline
12.10&sat&{\bf 0.53}&sat&11.15&5&816,017	\\
12.20&sat&{\bf 1.73}&sat&26.21&8&1,081,988	\\
12.30&unsat&334.11&unsat&{\bf 124.97}&200: 71$\rightarrow$68&358,731	\\
\hline
14.10&sat&{\bf 0.77}&sat&21.85&8&711,168	\\
14.20&sat&{\bf 3.25}&sat&21.07&23&155,345	\\
14.30&sat&26.33&sat&{\bf 3.36}&21&33,427	\\
\hline
16.10&sat&{\bf 0.50}&sat&12.06&6&435,640	\\
16.20&sat&{\bf 3.41}&sat&12.97&17&125,409	\\
16.30&unknown&2,732.93&unsat&{\bf 28.76}&200: 72$\rightarrow$64&45,918	\\
\hline
18.10&sat&{\bf 0.86}&sat&74.10&21&460,550	\\
18.20&unsat&677.83&unsat&{\bf 22.38}&200: 67$\rightarrow$56&30,431	\\
18.30&unsat&419.98&unsat&{\bf 37.89}&200: 71$\rightarrow$60&64,990	\\
\hline
20.10&sat&{\bf 1.39}&sat&26.58&14&118,920	\\
20.20&unknown&2,955.35&unsat&{\bf 28.26}&200: 58$\rightarrow$51&36,710	\\
20.30&unsat&58.99&unsat&{\bf 7.48}&200: 80$\rightarrow$51&10,049	\\
\hline
22.10&sat&12.42&sat&{\bf 6.61}&12&45,011	\\
22.20&unsat&1,720.84&unsat&{\bf 31.36}&200: 61$\rightarrow$57&22,140	\\
22.30&unsat&93.60&unsat&{\bf 9.67}&200: 75$\rightarrow$64&13,925	\\
\hline
24.10&sat&18.43&sat&{\bf 8.71}&18&35,563	\\
24.20&unknown&2,957.03&unsat&{\bf 16.12}&200: 65$\rightarrow$48&15,497	\\
24.30&unsat&103.70&unsat&{\bf 8.01}&200: 69$\rightarrow$58&7,033	\\
\hline
26.10&sat&{\bf 22.63}&sat&25.49&32&54,792	\\
26.20&unsat&1,931.69&unsat&{\bf 10.80}&200: 66$\rightarrow$52&10,848	\\
26.30&unsat&494.26&unsat&{\bf 5.04}&200: 75$\rightarrow$61&4,289	\\
\hline
28.10&sat&113.78&sat&{\bf 38.68}&48&49,115	\\
28.20&unsat&974.26&unsat&{\bf 23.96}&200: 71$\rightarrow$43&17,371	\\
28.30&unsat&457.82&unsat&{\bf 4.65}&200: 76$\rightarrow$60&3,799	\\
\hline
30.10&unknown&2,834.44&unsat&{\bf 91.84}&200: 50$\rightarrow$45&57,585	\\
30.20&unsat&1,007.06&unsat&{\bf 10.73}&200: 63$\rightarrow$56&10,095	\\
30.30&unsat&470.43&unsat&{\bf 3.27}&200: 83$\rightarrow$56&2,568	\\
\hline
32.10&sat&413.85&sat&{\bf 17.29}&43&19,008	\\
32.20&unsat&361.16&unsat&{\bf 11.26}&200: 59$\rightarrow$69&11,292	\\
32.30&unsat&148.69&unsat&{\bf 2.57}&200: 71$\rightarrow$72&2,202	\\
\hline
34.10&sat&61.41&sat&{\bf 7.97}&22&22,756	\\
34.20&unsat&192.02&unsat&{\bf 4.74}&200: 60$\rightarrow$53&3,982	\\
34.30&unsat&1,416.92&unsat&{\bf 2.88}&200: 75$\rightarrow$55&2,294	\\
\hline
36.10&unknown&3,316.51&unsat&{\bf 30.17}&200: 54$\rightarrow$54&18,460	\\
36.20&unsat&634.06&unsat&{\bf 6.98}&200: 53$\rightarrow$62&5,249	\\
36.30&unsat&107.73&unsat&{\bf 2.08}&200: 83$\rightarrow$59&1,740	\\
\hline
38.10&unknown&2,307.16&unsat&{\bf 18.69}&200: 59$\rightarrow$42&14,255	\\
38.20&unsat&235.11&unsat&{\bf 4.57}&200: 73$\rightarrow$55&3,821	\\
38.30&unsat&654.09&unsat&{\bf 2.44}&200: 79$\rightarrow$51&1,785	\\
\hline
\end{tabular}
\end{table}

\begin{table}[!th]\centering\small\setlength{\tabcolsep}{3pt}
\caption{Experimental test results for $k = 25$} \label{tbl:results-k-equal-25}
\begin{tabular}{|r|c|r||c|r|r|r|}
\cline{2-7}
  \multicolumn{1}{c}{} & \multicolumn{2}{|c||}{\bf Solver SAT4J} & \multicolumn{4}{c|}{\bf Solver FPT} \\
\hline
{\footnotesize Instance ID} &{\footnotesize Output} &{\footnotesize CPU Time\,(s)} &{\footnotesize Output} &{\footnotesize CPU Time\,(s)} &{\footnotesize \#Users} &{\footnotesize \#Patterns} \\
\hline
22.10&sat&{\bf 39.38}&sat&2,989.47&23&5,892,335	\\
22.20&sat&{\bf 107.45}&sat&2,985.02&75&1,413,105	\\
22.30&unsat&{\bf 380.64}&unsat&697.76&250: 84$\rightarrow$68&243,780	\\
\hline
24.10&sat&{\bf 29.38}&sat&1,317.98&14&4,547,403	\\
24.20&unknown&2,366.72&unsat&{\bf 1,041.61}&250: 89$\rightarrow$41&306,664	\\	
24.30&unsat&734.43&unsat&{\bf 530.52}&250: 87$\rightarrow$68&166,905	\\
\hline
26.10&unknown&{\bf 2,902.77}&sat&9,624.71&95&3,651,747	\\
26.20&unknown&2,629.04&unsat&{\bf 1,892.92}&250: 78$\rightarrow$47&447,506	\\
26.30&unsat&606.36&unsat&{\bf 343.73}&250: 86$\rightarrow$56&85,438	\\
\hline
28.10&sat&{\bf 359.67}&sat&3,200.61&44&1,914,685	\\
28.20&unknown&2,900.17&unsat&{\bf 1,190.27}&250: 86$\rightarrow$54&294,422	\\
28.30&unsat&{\bf 157.78}&unsat&259.69&250: 88$\rightarrow$61&62,806	\\
\hline
30.10&unknown&2,787.75&sat&{\bf 1,628.47}&29&1,354,688	\\
30.20&unknown&3,193.69&unsat&{\bf 1,367.66}&250: 74$\rightarrow$49&330,524	\\
30.30&unsat&753.05&unsat&{\bf 147.00}&250: 91$\rightarrow$65&37,563	\\
\hline
32.10&sat&{\bf 164.75}&sat&471.68&14&811,118	\\
32.20&unknown&2,294.51&unsat&{\bf 391.75}&250: 75$\rightarrow$55&101,370	\\
32.30&unsat&351.33&unsat&{\bf 142.21}&250: 89$\rightarrow$65&34,466	\\
\hline
34.10&unknown&3,510.65&unsat&{\bf 3,489.42}&250: 64$\rightarrow$40&800,535	\\
34.20&unknown&2,155.26&unsat&{\bf 400.66}&250: 83$\rightarrow$55&109,214	\\
34.30&unsat&576.72&unsat&{\bf 94.54}&250: 90$\rightarrow$67&19,242	\\
\hline
36.10&unknown&3,482.64&unsat&{\bf 1,960.90}&250: 65$\rightarrow$53&489,635	\\
36.20&unknown&2,935.06&unsat&{\bf 353.27}&250: 73$\rightarrow$50&89,912	\\
36.30&unsat&287.47&unsat&{\bf 68.07}&250: 97$\rightarrow$59&13,848	\\
\hline
\end{tabular}
\end{table}

For $k=20$ steps, Solver FPT successfully solved all $45$ test instances, but Solver SAT4J failed to reach a conclusive decision for $6$ of the instances ($13.3\%$), all $6$ being unsatisfiable (as determined by Solver FPT).
For small values of $b$ or $d$, corresponding to lightly-constrained instances, SAT4J usually performs better than Solver FPT. 
This is unsurprising, because the diversity $w$ of such instances is much higher, thereby increasing the running time of Algorithm~\ref{alg:pseudo-code}.
However, for larger values of $b$ or $d$ (where the diversity is much lower), the FPT algorithm clearly outperforms SAT4J. 
Also, Figure~\ref{fig:20} clearly shows that the running time of Solver SAT4J varies much more than that of Solver FPT, with the unsolved instances having running times larger than those of any solved instances. In contrast, Solver FPT shows a very stable time performance.
Solver SAT4J seems to be better to use when $b\le 10$, or $b\le 16$ and $d\le 20\%$, or $b\le 20$ and $d\le 10\%$.\\

%
%


For $k=25$ steps, Solver FPT reached a conclusive decision (satisfiable or unsatisfiable) for all the $24$ test instances.
In contrast, SAT4J failed to solve $11$ of the instances ($45.8\%$), two of which were reported to be satisfiable and nine of which were reported to be unsatisfiable by Solver FPT.
Again, for smaller values of $b$ or $d$, corresponding to lightly constrained instances, SAT4J generally performs better than Solver FPT. 
However, for larger values of $b$ or $d$, SAT4J starts to fail quite often and is not able to provide a solution. At the same time, our FPT algorithm clearly starts to outperform SAT4J because the diversity $w$ of such instances is much lower.
Figure~\ref{fig:25} shows that, for $25$ steps, the running time of Solver SAT4J varies similarly to the running time of Solver FPT, with an outlier instance at $b=26$ and $d=10\%$ (satisfiable).  
However, SAT4J is unable to solve almost half of the instances, with the unsolved instances again having running times higher than any solved instances.
Solver SAT4J seems to be better to use when $b\le 22$, or $b\le 24$ and $d\le 10\%$, or when an instance is highly suspected to be satisfiable.

\subsubsection{Summary}

Table~\ref{tbl:results-different-k} above presents summary statistics for the experiments overall. 
From Tables~~\ref{tbl:results-k-equal-20} and \ref{tbl:results-k-equal-25} and the low running time variance for $k=15$, it can be observed that the average running times of Solver FPT are of a similar order of magnitude, whether the instances are satisfiable or unsatisfiable.
The set of running times for Solver FPT in Table~~\ref{tbl:results-k-equal-20}, whether the instance is satisfiable or unsatisfiable, also has relatively low variance. 
In contrast, the mean running times of SAT4J vary significantly depending on whether the instance is satisfiable or not.
As the number $k$ of steps increases, SAT4J fails more frequently, and is unable to reach a conclusive decision for almost half of the instances when $k = 25$.
This is unsurprising, given that the number of PB SAT variables will grow quadratically as $k$ and $n=10k$ increase.
However, in time performance for satisfiable instances, the picture is often more favourable to SAT4J: this may be explained by some heuristics deployed to solve relatively easy satisfiable instances.
Overall, for larger values of $k$, the average run-time advantage of the FPT algorithm over SAT4J decreases, but the relative number of instances solved by SAT4J decreases as well.

The tables also exhibit the expected correlation between  the running time of our FPT algorithm and two numbers: the number of patterns generated by the algorithm and the number of users considered, which, in turn, is related to the number of constraints and constraint density.
For well-constrained instances, the FPT algorithm has to consider far fewer patterns, and this more than offsets the fact that we may have to consider every user (for those cases that are unsatisfiable).


It is interesting to note the way in which the mean running time $\hat{t}$ varies with the number of steps.
In particular, $\hat{t}$ for our algorithm grows exponentially with $k$ (with strong correlation between $k$ and $\log \hat{t}$), which is consistent with the theoretical running time of our algorithm ($O^*(2^{k \log k})$).
The running time of SAT4J is also dependent on $k$, with a strong correlation between $k$ and $\log \hat{t}$, which is consistent with the fact that there are $O(n^k)$ possible plans to consider.
However, it is clear that the running time of SAT4J is also more dependent on the number of variables (determined by the number of users, authorizations, and constraints), than it is on $k$, unlike the running time of our algorithm.


\section{Concluding Remarks}\label{sec:conclusion}

In this paper, we described our implementation of an FPT algorithm designed to solve a specific NP-hard problem known as the workflow satisfiability problem (WSP) for user-independent counting constraints.
In theory, there exists an algorithm that can solve the WSP for user-independent constraints in time $O^*(2^{k\log k})$ in the worst case.
However, the WSP is a practical problem with applications in the design of workflows and the design of access control mechanisms for workflow systems~\cite{CrGu13}.
Thus, it is essential to demonstrate that theoretical advantages can be realised in practice.

Accordingly, we have developed several implementations using the generic FPT algorithm as a starting point.
In developing the implementations, it became apparent that several application-specific heuristic improvements could be made.
In particular, we developed specific types of propagation and pruning techniques for counting at-most-$r$ and at-least-$r$ constraints.
Following general techniques described in Section~\ref{sec:solving-wsp:fpt-algorithm} and in \cite{CoCrGaGuJo13},
it should be possible to generalize and implement efficiently most of the ideas used in Algorithm~\ref{alg:pseudo-code} to solve the WSP with other types of user-independent constraints.

We compared the performance of our algorithm with that of SAT4J---an ``off-the-shelf'' PB SAT solver.
In order to perform this comparison, we extended Wang and  Li's encoding of the WSP as a pseudo-Boolean satisfiability problem \cite{WaLi10}.
The results of our experiments suggest that our algorithm does, indeed, have an advantage over SAT4J when solving the WSP, although this advantage does not extend to lightly constrained instances of the problem.
The results also suggest that those advantages could be attributed to the structure of our algorithm, with its focus on the small parameter (in this case, the number of workflow steps).

The encodings of plans presented in this paper grouped plans together based on which steps they assigned to the same user.
For counting constraints, it would also be possible to group plans together based on how many users are assigned to the steps in each constraint.
It may be worth investigating algorithms based on this encoding of plans in future.

We plan to continue working on algorithm engineering for the WSP. In particular, we plan to continue developing  ideas presented in this paper and in \cite{CoCrGaGuJo13} to develop efficient implementations and modified versions of this FPT algorithm. We hope to obtain a more efficient implementation than the one presented in this paper.
We also plan to try different experimental setups. For example, in this paper, we have used a uniform random distribution of authorizations to users with an upper bound at 50\% of the number of steps for which any one user can be authorized. In some practical situations, a few users are authorized for many more steps than others.
We have only considered counting constraints, rather than a range of user-independent constraints.
In some ways, imposing these constraints enables us to make meaningful comparisons between the two different algorithmic approaches. However, we would like to undertake more extensive study and testing to confirm that the initial results obtained for this particular family of WSP instances can be extended to other types of WSP.

Results and ideas presented in this paper can serve as a benchmark for further developments in algorithm engineering to solve the workflow satisfiability problem with user-independent constraints and to design their experimental testing.

\paragraph{Acknowledgments}
This research was supported by EPSRC grant EP/K005162/1.\\ 
We are very grateful to the referees for several useful comments and suggestions and to
Daniel Karapetyan for 
several helpful discussions.


\begin{thebibliography}{10}
\providecommand{\url}[1]{\texttt{#1}}
\providecommand{\urlprefix}{URL }

\bibitem{ANSI04}
American National Standards Institute. \emph{ANSI INCITS 359-2004 for Role Based Access Control}, 2004.

\bibitem{BaBuKa14}
Basin, D.A., Burri, S.J., Karjoth, G.: 
Obstruction-free authorization enforcement: Aligning security and business objectives. 
\emph{Journal of Computer Security} 22(5), 661-698 (2014).

\bibitem{BeTa10}
Berend, D., Tassa, T.: Improved bounds on {Bell} numbers and on moments of sums
  of random variables. Probability and Mathematical Statistics  30(2),
  185--205 (2010)

\bibitem{BeFeAt99}
Bertino, E., Ferrari, E., Atluri, V.: The specification and enforcement of
  authorization constraints in workflow management systems. ACM Trans. Inf.
  Syst. Secur.  2(1),  65--104 (1999)

\bibitem{ChKl10}
Chimani, M., Klein, K.: Algorithm engineering: Concepts and practice. In:
  Bartz-Beielstein, T., Chiarandini, M., Paquete, L., Preuss, M. (eds.)
  Experimental methods for the analysis of optimization algorithms, pp.
  131--158 (2010)

\bibitem{FAW2014}
Cohen, D., Crampton, J., Gagarin, A., Gutin, G., Jones, M.: 
Engineering algorithms for workflow satisfiability problem with user-independent constraints. 
Proc. 8th International Frontiers of Algorithmics Workshop (FAW 2014), J. Chen, J.E. Hopcroft, and J. Wang (Eds.), 2014, LNCS 8497, Springer, pp. 48-59.

\bibitem{CoCrGaGuJo13}
Cohen, D., Crampton, J., Gagarin, A., Gutin, G., Jones, M.: 
Iterative plan construction for the workflow satisfiability problem. 
Journal of Artificial Intelligence Research 51, 555--577  (2014)

\bibitem{Cr05}
Crampton, J.: A reference monitor for workflow systems with constrained task
  execution. In: Ferrari, E., Ahn, G.J. (eds.) SACMAT. pp. 38--47. ACM (2005)

\bibitem{CrGu13}
Crampton, J., Gutin, G.: Constraint expressions and workflow satisfiability.
  In: Conti, M., Vaidya, J., Schaad, A. (eds.) SACMAT. pp. 73--84. ACM (2013)

\bibitem{CrGuYe13}
Crampton, J., Gutin, G., Yeo, A.: On the parameterized complexity and
  kernelization of the workflow satisfiability problem. ACM Trans. Inf. Syst.
  Secur.  16(1), ~4 (2013)

\bibitem{DowneyFellows99}
Downey, R.G., Fellows, M.R.: Fundamentals of Parameterized Complexity. Springer Verlag (2013)

\bibitem{Du64}
Durstenfeld, R.: Algorithm 235: Random permutation. Communications of the ACM
  7(7),  420 (1964)

\bibitem{FiYa49}
Fisher, R.A., Yates, F.: Statistical tables for biological, agricultural and
  medical research. Oliver and Boyd, third edn. (1948)

\bibitem{FlumGrohe06}
Flum, J., Grohe, M.: Parameterized Complexity Theory. Springer Verlag (2006)

\bibitem{GlGaFe98}
Gligor, V.D., Gavrila, S.I., Ferraiolo, D.F: 
On the formal definition of separation-of-duty policies and their composition. 
In \emph{IEEE Symposium on Security and Privacy}, pp.\,172--183. IEEE Computer Society, 1998.


\bibitem{BePa10}
Le~Berre, D., Parrain, A.: The {SAT4J} library, release 2.2. J. Satisf. Bool.
  Model. Comput.  7,  59--64 (2010)

\bibitem{MyKo11}
Myrvold, W., Kocay, W.: Errors in graph embedding algorithms. J. Comput. Syst.
  Sci.  77(2),  430--438 (2011)

\bibitem{Niedermeier06}
Niedermeier, R.: Invitation to Fixed-Parameter Algorithms. Oxford U. Press
  (2006)

\bibitem{ReNiDe77}
Reingold, E.M., Nievergelt, J., Deo, N.: Combinatorial algorithms: Theory and
  practice. Prentice Hall (1977)
  
\bibitem{ScSpWe05} Schaad, A., Spadone, P., Weichsel, H.:
  A case study of separation of duty properties in the context of the Austrian ``eLaw" process.
   Proc. the 2005 {ACM} Symposium on Applied Computing (SAC 2005), 
  1328--1332 (2005)
  
\bibitem{ScLoSo06} Schaad, A., Lotz, V., Sohr, K.: 
  A model-checking approach to analysing organisational controls in a loan origination process. In: Ferraiolo, D.F., Ray, I. (eds)
  SACMAT. pp. 139--149. ACM (2006)
  
\bibitem{WaLi10}
Wang, Q., Li, N.: Satisfiability and resiliency in workflow authorization
  systems. ACM Trans. Inf. Syst. Secur.  13(4), ~40 (2010)

\bibitem{WoSch07}
Wolter, C., Schaad, A.:
Modeling of task-based authorization constraints in BPMN. 
In G. Alonso, P. Dadam, and M. Rosemann (eds), BPM, LNCS 4714, pp.\,64-79. Springer, 2007.


\end{thebibliography}

\end{document}